\documentclass[conference]{IEEEtran}
\IEEEoverridecommandlockouts
\usepackage{cite}
\usepackage{amsmath,amssymb,amsfonts}
\usepackage{algorithmic}
\usepackage{graphicx}
\usepackage{textcomp}
\usepackage{xcolor}
\usepackage{hyperref}
\def\BibTeX{{\rm B\kern-.05em{\sc i\kern-.025em b}\kern-.08em
    T\kern-.1667em\lower.7ex\hbox{E}\kern-.125emX}}
\begin{document}

\title{Lost in Transition: The Struggle of Women Returning to Software Engineering Research after Career Breaks}

\author{\IEEEauthorblockN{Shalini Chakraborty}
\IEEEauthorblockA{University of Bayreuth \\
shalini.chakraborty@uni-bayreuth.de}
\and
\IEEEauthorblockN{Sebastian Baltes}
\IEEEauthorblockA{University of Bayreuth \\
sebastian.baltes@uni-bayreuth.de}

}
\maketitle

\begin{abstract}
The IT industry provides supportive pathways such as returnship programs, coding boot camps, and buddy systems for women re-entering their job after a career break.
Academia, however, offers limited opportunities to motivate women to return. 
We propose a diverse multicultural research project investigating the challenges faced by women with software engineering (SE) backgrounds re-entering academia or related research roles after a career break.
Career disruptions due to pregnancy, immigration status, or lack of flexible work options can significantly impact women's career progress, creating barriers for returning as lecturers, professors, or senior researchers.
Although many companies promote gender diversity policies, such measures are less prominent and often under-recognized within academic institutions.
Our goal is to explore the specific challenges women encounter when re-entering academic roles compared to industry roles; to understand the institutional perspective, including a comparative analysis of existing policies and opportunities in different countries for women to return to the field; and finally, to provide recommendations that support transparent hiring practices. 
The research project will be carried out in multiple universities and in multiple countries to capture the diverse challenges and policies that vary by location. 
\end{abstract}


\section{Introduction}
Women in software engineering (SE) who wish to return to academia, whether as lecturers, professors, or in other research roles, often navigate distinct and layered challenges.
Those challenges direct them in taking a break in the first place, but also affect their attempts to re-enter the field.
For many women, career breaks come from pregnancy~\cite{marbury1984work}, the need to balance caregiving responsibilities~\cite{moen1994women, lee2015more}, manage relocations due to family~\cite{collier1994gender} or immigration, or deal with health and personal matters. Academic institutions typically offer limited flexibility and structured support, making it difficult for women to stay in or quickly return to research or teaching roles after interruptions. Unlike the IT sector, where returnship programs, buddy systems, and upskilling boot camps are common~\cite{holtzblatt2022retaining}, academia often lacks similar resources. 
Many academic institutions lack gender diversity policies for returnees, creating hiring gaps, whereas the IT industry offers programs to support re-entry after a career break.
Big tech companies such as Microsoft (Springboard\footnote{\href{https://news.microsoft.com/en-in/features/springboard-way-sabbatical/}{https://news.microsoft.com/en-in/features/springboard-way-sabbatical/}}) and Google (Next Innings\footnote{\href{https://www.jobsforher.com/events/next-innings-returnship-program-for-women-engineers/1074}{https://www.jobsforher.com/events/next-innings-returnship-program...}}) have each launched returnship programs that provide women opportunities for getting back into the IT sector. This contrast means that women returning to academia must navigate not only the demands of their field but also institutional and structural barriers that are far less prevalent in the IT industry, with which academic SE departments indirectly compete when hiring faculty.

\section{Women returning to academia vs industry}
Despite a constant demand for university lecturers, a significant population is struggling to return to academic roles. 
Building a sustainable and balanced society requires diverse representation in education, research, and innovation. However, women remain significantly underrepresented in SE research~\cite{Happe2022frustration}. In Germany, women accounted for only 29\% of the research and development workforce in 2021\footnote{\href{https://www.destatis.de/Europa/EN/Topic/Science-technology-digital-society/FemaleResearchers.html}{https://www.destatis.de/Europa/EN/Topic/Science-technology-digital...}}. 
This trend is reflected globally, with similarly low numbers in most countries. In a study, Mattauch et al. examined 19 major computer science conferences over six years (2014-2019) and found that, on average, less than 10\% of the contributors were women, underscoring the persistent gender gap in fields critical to innovation and technological progress~\cite{mattauch2020bibliometric}.
Studies also discuss women returning to industry~\cite{herman2015rebooting,panteli2012community} but there is little to no literature on women returning to academia. If we want to overcome gender bias and build a diverse SE practice, we must explore both industry and academia. 
The first obstacle is that, unlike industry, the hiring policies in academia are not well advertised in terms of diversity and inclusion. Some universities include statements such as \emph{we encourage underrepresented genders to apply for the job}. However, the question remains: How much does this affect women returning to research? Conferences such as \textbf{ACM WomENcourage}\footnote{\href{https://womencourage.acm.org/2025/}{https://womencourage.acm.org/2025/}} have been promoting women's contribution to computer science, but women are still underrepresented in academia, especially in SE. In their study, Breukelen et al. highlight the challenges faced by veteran women in SE practice and their strategies to survive~\cite{van2023still}. 
However, we are not aware of any study that recognizes the experiences of veteran women in SE research. Staying in SE research can be just as challenging, if not more so, than remaining in SE practice, yet this aspect remains underexplored. 
%
\section{Challenges}
For women attempting to resume academic careers, the absence of clear, supportive re-entry pathways presents a major challenge. To collect data for understanding challenges to return, we have divided them into the following categories, which are based on existing literature on returnship programs for IT and our own hypotheses, as there is almost no literature on women returning to academia. 
\begin{itemize}
    \item \textbf{Cultural factors}: Women immigrating to developed countries often had to pause their careers due to cultural clashes and family pressure. Such factors linger and make it hard for them to return to academia. Language can also play a major role, because a teacher or professor needs strong communication skills. However, non-linguistic challenges such as structural barriers or responsibilities at home can be just as harmful. In the book \emph{Immigrant Women Tell Their Stories}~\cite{berger2013immigrant}, the author describes in the participant's own words how difficult it is for women to focus on their careers due to cultural reasons that are unknown in most developed countries. 
    \item \textbf{Organizational resistance}: In academic institutions, hiring policies typically emphasize recent research experience, meaning a gap in one’s CV can significantly influence hiring decisions. As Mavriplis et al. (2010) describe, the \emph{``academic model for advancement in scientific disciplines includes a preference for a lock-step career progression from undergraduate to graduate education, to a postdoctoral position and then to an academic position with continuous employment, especially as relates to tenure-track positions, and large amounts of contact time especially in lab-based disciplines, accompanied by an expectation that one's career is ``made" in one's 30s"}~\cite{mavriplis2010mind}. Due to this rigid career structure in academic STEM fields, part-time employment is rarely viewed as a viable path toward advancement and is often perceived as a sign that a candidate is ``not serious" about their career.
    \item \textbf{Lack of tech support}: Academic positions, such as professorships, often lack the technical and collaborative support available in typical SE roles. In the IT industry, employees commonly rely on advanced technical tools, collaborative software, and team-based environments that help maintain and develop skills even after a career break. Additionally, structured upskilling resources and continuous training in new technologies allow software engineers to stay competitive and ease back into their roles. In contrast, academia, particularly in teaching and research, is largely self-driven~\cite{bailey1999academics}, where professors are expected to stay independently updated on research trends and publication standards. A background in full stack or frontend development only helps you a little when you want to return to academic roles. Unlike the tech industry, academia offers few tools or formal resources to help returning faculty bridge the gap, often leaving them to navigate re-entry on their own.
    %
    \item \textbf{Lack of awareness}: A significant challenge for women returning to academic roles is the lack of awareness and visibility surrounding diversity and re-entry programs. Unlike industry, where diversity initiatives, despite the backlash, have become mainstream, academia still lacks prominent, accessible programs tailored to support women returning after career breaks. In the Nordics, the unemployment offices work closely with industry but not with universities. Academic institutions may be unaware of the unique obstacles faced by returning women, such as re-engaging with research, managing publication gaps, and adjusting to evolving academic standards. 
\end{itemize}

In addition to the four challenges mentioned above, it has been established in previous studies that gender bias and ageism prevent women from returning to work~\cite{wang2019implicit,liang2024controlled,o2025assessing}. 

\section{Project Proposal}
\textbf{We need a thorough investigation to understand the struggle of women returning to SE academia}.  Policies and legislation differ between countries, and we understand that this can also impact return journeys. Therefore, keeping the vision in mind, we are proposing a diverse, multicultural research project that is currently being planned to be executed in three different countries: Germany, Canada, and Brazil. We chose these three countries because they offer diverse cultural perspectives and have significant immigrant populations. Once we obtain our initial results, we can expand our study to include additional countries. Our current focus is on comparing the situations of returning women between industry and academia in the selected countries. A comparative analysis between the two sectors could help us understand what is lacking in academia and which industry policies academia might adopt. Note that the project is in its early stages. Thus, the scope is open to exploration. We aim to interview women at various academic stages, from postdocs to assistant professors, to explore their motivations, challenges, and experiences in returning to academia versus industry. 
In the next step, we want to explore existing policies around women returning to academia in those three countries to understand if any significant difference exists compared to the industry, which is limiting women's return to academia. 

\section{Conclusion}
In conclusion, addressing the challenges faced by women returning to academia is essential to foster inclusive and diverse SE practices. Our research, spanning multiple countries, seeks to capture the varied experiences of women re-entering academia, identifying both the personal and structural obstacles they face. We aim to identify the gaps between academia and industry and offer practical recommendations for transparent hiring practices that encourage women's return. We welcome feedback and invite researchers to collaborate, as we need more diverse data to support a balanced and equitable academic environment for future generations.

\newpage

\bibliographystyle{IEEEtran}
\bibliography{IEEEabrv,refs}

\end{document}